\documentclass[aps,twocolumn,superscriptaddress,floatfix]{revtex4-1}

\usepackage[pdftex]{graphicx} % Allows for eps images
\usepackage{epsfig}
\usepackage{amsmath} % AMS Math Package
\usepackage{amsthm} % Theorem Formatting
\usepackage{amssymb}	% Math symbols such as \mathbb
\usepackage{physics}

\usepackage{hyperref} %clickable references.
\hypersetup{
    colorlinks,
    citecolor=blue,
    filecolor=black,
    linkcolor=red,
    urlcolor=blue
}

\newcommand{\dmin}{D^2_{\min}}

\newcommand{\delmin}{\Delta^2_{\min}}

\usepackage{soul}

\usepackage[english]{babel}
\usepackage{blindtext}
\usepackage{makecell}
\usepackage{tikz}

\begin{document}

\bibliographystyle{unsrt}

\title{Correlation of plastic events with local structure in jammed packings across spatial dimensions}
\author{Sean A. Ridout}
\author{Jason W. Rocks}
\author{Andrea J. Liu}
\affiliation{Department of Physics and Astronomy, University of Pennsylvania, Philadelphia, PA 19104, USA}

\begin{abstract}
  In jammed packings, it is usually thought that local structure only plays a significant role in specific regimes. The standard deviation of the relative excess coordination, $\sigma_Z/ Z_\mathrm{c}$, decays like $1/\sqrt{d}$, so that local structure should play no role in high spatial dimensions. Furthermore, in any fixed dimension $d \geq 2$, there are diverging length scales as the pressure vanishes approaching the unjamming transition, again suggesting that local structure should not be sufficient to describe response. Here we challenge the assumption that local structure does not matter in these cases. In simulations of jammed packings under athermal, quasistatic shear, we use machine learning to identify a local structural variable, softness, that correlates with rearrangements in dimensions $d=2$ to $d=5$. We find that softness - and even just the coordination number $Z$ -  are quite predictive of rearrangements over a wide range of pressures, all the way down to unjamming, in all $d$ studied. This result provides direct evidence that local structure can play a role in higher spatial dimensions.
\end{abstract}
\maketitle

%\begin{figure}[t!]
%\centering
%\includegraphics[width=0.95\linewidth]{example.pdf}
%\caption{}
%\label{fig:filtration}
%\end{figure}

% \comment{Alternate title: Local structure as a predictor of plastic events in jammed packings across spatial dimensions

% Correlation of plastic events with local structre in}

Supercooled liquids, glasses and jammed packings can relax via rearrangements, in which particles experience large, sudden displacements, due to thermal fluctuations or mechanical loading. In such systems in $d=2$ and $d=3$, where $d$ is the spatial dimension, the details of local structure play an important role, an idea supported by numerous studies that correlate local structure with rearrangements~\cite{Reichman2008,Reichman2009,Manning2011,SamPRX,barrat,Gartner2016,Zylberg2017,malins,Tanaka2018,Richard2020,tetrahedrality,Bapst2020,boattini2020autonomously,Viitanen2020}. In particular, machine learning has been used to identify a highly predictive combination of local structural  descriptors, named ``softness,'' which provides insight into the underlying physics~\cite{Cubuk2015, Schoenholz2016, Sussman2017, Sharp2018, Harrington2018, Ma2019, Landes2020,Cubuk2020,Tah2020,Zhang2020, Rocks2019, Richard2020}. For example, in thermal systems, each value of this continuous structural variable is associated with a characteristic free energy barrier to rearrangements that decreases with increasing softness~\cite{Schoenholz2016, Sharp2018,Ma2019,Landes2020,Cubuk2020,Tah2020,Zhang2020}. 

On the other hand, a successful analytical approach to these systems has been found in the form of a mean-field theory. These mean-field calculations, valid in the limit $d=\infty$, seem not to include details of local structure, and activated processes - which we associate with softness - are absent because their free energy barriers become infinite. In spite of these simplifications, these mean-field models predict many glass transition phenomena~\cite{Charbonneau2017} and Gardner transition phenomena~\cite{Berthier2019} with qualitative accuracy, even in $d=2,3$, and quantitatively predict aspects of jamming criticality in $d=2,3$~\cite{Charbonneau2017}. %There is evidence -- from theory, simulation and experiment --  that phase transitions predicted by the mean-field theory are present in dimensions two and higher~\cite{ Charbonneau2017b, Berthier2016, Jin2017, Seguin2016}.

The successes of these two frameworks seem to be in tension with one another; if the details of the local structure are empirically important in low dimensions, how can mean-field theories which ignore them be so successful? One pathway to reconciling these facts is to look for a smooth transition to mean-field-like behaviour as the dimension is raised. It is useful to address this question in the context of athermal, quasistatic shear to avoid the existence of multiple relevant time scales governing rearrangements that are inevitable at nonzero temperatures $T$. At $T=0$, there is reason to believe that localized rearrangements correlated with local structural variables give way with increasing $d$ to delocalized rearrangements that are poorly correlated with local structure. At $T=0$, rattlers and bucklers, two types of local structures associated with instabilities (rearrangements) as well as deviation from mean field predictions in jammed packings, grow exponentially more rare with increasing $d$~\cite{Charbonneau2012,Charbonneau2015}. Furthermore, as $d$ rises, a larger fraction of the low-frequency vibrational modes become extended, and the density of vibrational states smoothly approaches the form expected from mean field theory, although in any finite dimension there still exist some quasilocalized modes at sufficiently low frequency \cite{Charbonneau2016, Shimada2019}. 
These facts seem to imply that local structure does matter in low dimensions, but becomes less and less relevant as $d$ increases, smoothly converging to some ``mean-field-like'' dynamics analogously to the density of states and distribution of forces.

In light of these ideas, this paper quantifies directly the importance of local structure as a function of spatial dimension. Here we examine the prediction accuracy of machine-learned softness in jammed soft sphere packings under athermal, quasistatic shear, in dimensions $d=2$ to $d=5$. In contrast to the predominant belief that local structure only matters in low dimensions, we find that as $d$ increases, local structure as quantified by softness becomes \textit{more} predictive of rearrangements. Moreover, we find that softness increasingly coincides with simply the local number of interacting neighbors, or coordination number, $Z$. This result suggests that the distribution of $Z$ values plays an important role and that $Z$ could be a valuable indicator of mobility even in the mean field limit.

%\section{Simulation methods}

We prepare bi-disperse jammed packings of Hertzian particles at a range of pressures in each spatial dimension. We study $N=4096$ particles in $2d$,  $N=8192$ in $3d$ and $4d$, and $N=16384$ in $5d$.  Initially, particles are assigned random positions ($T=\infty$), before quenching to a energy minimum, producing a force-balanced, zero-temperature (jammed) state. 
The system is sheared quasistatically by applying small strain steps ($\delta \gamma = 10^{-4}$) and re-minimizing the energy $E$ with respect to $\left\{ x_i \right\}$ at each step. When the pressure $P$ is large, it is relatively insensitive to shear ($\frac{1}{P} \pdv{P}{\gamma} \approx 0$), allowing us to minimize with a fixed box volume $V$.
However, when $P$ is small and the system is finite (and thus slightly anisotropic), the pressure can fluctuate substantially with shear, so we fix $P$ by minimizing the enthalpy, $H = E + PV$, with respect to both $\left\{ x_i \right\}$ and volume.When a plastic event is detected in the form of a drop in the shear stress, the resulting state is discarded, and the stress drop is approached with a shear step $\delta \gamma / 2$ \cite{vanHecke2014, Morse2020}. We repeat this procedure until the stress drop is approached with a shear step of $\delta \gamma = 10^{-12}$, at which point the main numerical limitation to realizing true quasistatic shear is the force tolerance in the energy minimization algorithm.  

%\section{Training softness and testing its effectiveness}

Next, we quantify local structure, by finding a combination of local structural descriptors captures each particle's propensity to rearrange. In previous studies, populations of rearranging and non-rearranging particles have been identified from a collection of configurations and then used to train classifier to sort particles into these two groups~\cite{Cubuk2015}. 
Here we instead utilize a more conceptually straightforward approach in which linear regression is used to identify structures that correlate with local fluctuations in the displacement field~\cite{Rocks2019}. 
In addition to providing more concise sets of structural descriptors, this method has the added advantage of requiring far less data to construct a high quality training set as it can use all particles in a data set, rather than a small subset of examples.

In this approach, we describe the local structure around each particle by counting the number of contacts and gaps at each distance in a triangulation of the packing. 
Once the Delaunay triangulation of the packing is constructed, each particle $j$ in the triangulation is assigned a discrete distance $d_{ij}$ relative to particle $i$ such that $d_{ii}=0$, $d_{ij}=1$ for neighbours, $d_{ij}=2$ for particles which share a neighbor, etc. 
Each edge $\left(j, k\right)$ in the triangulation is either a contact between particles or a gap and may be assigned a distance from particle $i$, defined as $d_{i, \left(j,k\right)} = d_{ij} + d_{jk}$.
The number of gaps and contacts at each distance up to $d_{i, \left(j,k\right)}=8$, together with the particle radius, form a set of 17 structural descriptors that we use to encode the local structural environment around each particle.

One complicating factor is that quasistatically sheared jammed packings exhibit crackling noise in which one localized rearrangement triggers a second, which triggers a third, and so forth, producing an avalanche at each single stress drop or instability~\cite{Tewari1999}. 
To focus on localized rearrangements rather than avalanches, we follow what has become standard practice~\cite{Maloney2006,Manning2011,Richard2020}, designing our structural variable to predict the first rearrangement in each avalanche.
We capture this initial rearrangement by computing the lowest eigenvector of the dynamical matrix immediately before the stress drops. 
This eigenvector corresponds to the normal mode whose frequency vanishes at the onset of the rearrangement, signaling the instability. 
Using this eigenvector, we compute the quantity $\dmin$ for each particle, measuring its non-affine displacement relative to its neighbours~\cite{Falk1998,Maloney2006}. 

\begin{figure}[t!]
  \centering
  \includegraphics[width=\linewidth]{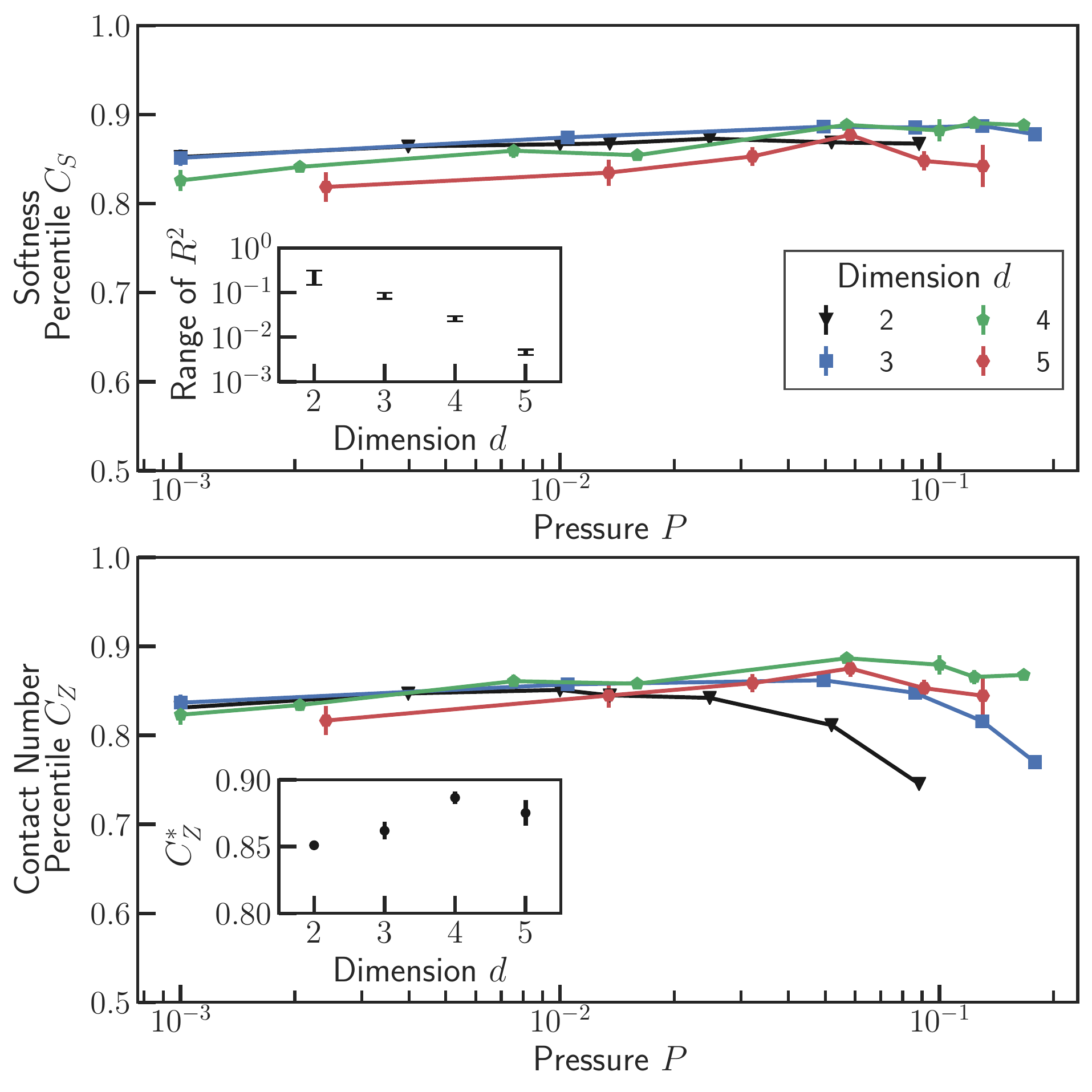}
  \caption{The $S$ percentile $C_S$ (A) and $-Z$ percentile $C_Z$ (B) of the particle with the highest $\delmin$ in a plastic event, as a function of dimension $d$ and pressure $P$ ($d=2$ black triangles, $d=3$ blue squares, $d=4$  green pentagons, $d=5$  red hexagons). $C_Z$ clearly peaks at a finite pressure, and decays at large pressure. $C_S$ appears to have a weak pressure dependence which is difficult to resolve in our data. Except at high pressure, $Z$ performs almost as well as $S$. Error bars are uncertainty in mean value. Inset to (A) $R^2$ for fitting of $S$ to the local displacement fluctuations $\delmin$. Error bars show full range over all pressures; all data are shown in the Supporting Information~\cite{Supplement}.  Although the rearranging particle is predicted equally well in higher $d$, the correlation to the rescaled mobility field used to fit $S$ is actually much worse in higher $d$,. Inset to (B): Peak value over all pressures $C^*_Z$, showing little dimension dependence. }
  \label{fig:monster_C_fig}
  \end{figure}

Because the system is solid, a rearranging particle exerts a long-ranged strain field on other particles in the system, causing a power-law decay with distance of $\dmin$ averaged over directions. A particle near the rearrangement core will move more than one far from it, regardless of whether their respective structures predispose them to mobility or to rigidity. To account for this effect, we rescale $\dmin$ for each particle by dividing by the average $\dmin$ of its neighbors~\cite{Rocks2019}; we denote this quantity as $\delmin$ and expect that it should better correlate with local structure than $\dmin$.

Next, we perform linear regression to identify a linear combination of our structural descriptors, the ``softness'' $S$, which best correlates with $\delmin$. Although the correlation coefficient $R^2$ is not high, and becomes weaker in higher dimensions, we find that the prediction accuracy on rearranging particles is just as high as previous classification-based approaches~\cite{Rocks2019}.
This quantification of prediction accuracy is superior to comparing classification accuracies on a training set or regression accuracies because these numbers change if the definition of the training set is changed, even if the resulting softness $S$ does not change. An additional issue is that, as described above, our training variable $\delmin$ does not even directly correspond to rearranging and non-rearranging particles so it does not make sense to calculate the prediction accuracy for rearranging particles. An alternative approach might be to set a threshold on $\dmin$ so that particles above this threshold are ``rearranging'', and then to ask which fraction of these particles have high softness~\cite{Zhang2020}. However, this means of quantifying the prediction accuracy of softness is problematic in our case because a different threshold $\dmin$ value would have to be selected in different dimensions and pressures, making meaningful comparisons across pressures and dimensionalities difficult.

Once training is complete, each particle in the system is assigned a value of $S$. 
To evaluate the predictive power of $S$, or any structural indicator of rearrangements, a standard approach is to check whether the particle most associated with the center of each rearrangement also has a high value of the softness $S$~\cite{Patinet2016,Richard2020}. 
We identify this particle as that with the maximum value of  $\dmin$ in the critical mode. 
We them measure the average percentile  of $S$ for rearranging particles, which we call $C_S$, as our measure of correlation between structure and dynamics.
 A value of $C=1$ corresponds to perfect prediction accuracy while $C=0.5$ corresponds to random guessing. 
 
In Fig.~\ref{fig:monster_C_fig}(A), we report $C_S$ for packings in all dimensions as a function of pressure $P$.
We compute error bars as an estimate of the uncertainty in our measures due to sampling error calculated via cross-validation~\cite{Rocks2019}.
We find that within uncertainty the predictive power of $S$ does not strongly depend on either pressure nor dimension.
At lower pressures in $d=5$, $C_S$ may be slightly less predictive, but it is otherwise difficult to see any significant trends.

Although Fig.~\ref{fig:monster_C_fig} shows that the local structure can predict the location of the maximum of $\dmin$ equally well in all dimensions studied, we note that softness is actually less predictive of the local fluctuations in $\dmin$ which we use to train $S$ at higher $d$. The $R^2$ of the linear regression, shown in the inset of the first panel, decays dramatically $2$ and $5$. Thus, in higher dimensions, the local fluctuations in $\dmin$ are actually very poorly correlated with the local structure, but what little correlation remains is enough to identify a structural variable which correlates strongly with the rearrangement itself.

Furthermore, we find that most of the predictive power of $C_S$ is determined by the number of contacts at distance $d_{i, (j,k)}=1$, i.e. the coordination number $Z$ of each particle.
As a comparison, we compute the predictive power of the coordination number $Z$ by computing the average percentile of $-Z$ , which we denote as $C_Z$.
In Fig.~\ref{fig:monster_C_fig}(B), we report $C_Z$ for the same set of particles used to measure $C_S$.
Except at high pressures, we see that $C_Z$ is comparable to $C_S$; in $4d$ and $5d$ they are equal within the uncertainty. Thus, the coordination in fact contains most of the local structural information we are capable of finding using our method -- in high dimension it contains the \textit{only} available information.

In order to more easily compare the properties of our structure and dynamics across dimensions, 
we pick a single pressure to represent each dimension.
Comparing systems at the same pressure is not necessarily meaningful; pressure has different units in different spatial dimensions.
Alternative control parameters include the difference in packing fraction from the jamming transition $\Delta \phi$, the excess coordination number $\Delta Z$ needed for rigidity, and the ``prestress'' $e \sim \Delta Z / d$ which has recently been suggested as an appropriate control parameter when comparing between dimensions \cite{Shimada2019}.
However, we find that none of these choices results in a clean collapse of the curves for $C_Z$ in Fig.~\ref{fig:monster_C_fig}(B). 
Therefore, whenever we need to compare a property of $S$ or $Z$ across spatial dimension, 
we choose the pressure $P^*(d)$ in each dimension which maximizes $C_Z$.
In the inset of Fig.~\ref{fig:monster_C_fig}(B), we report the predictive power of $Z$ at $P^*(d)$, which we denote  $C^*_Z$. 
We find that $C^*_Z$ appears to be approximately constant between $d=3,4,5$ within sampling error.

Surprisingly, neither $C_S$ nor $C_Z$ shows a strong decrease at low pressure. 
It is known that various length scales diverge at the jamming transition, including the length scale corresponding to spatial correlations of $Z$, as well as the size of a localized ``core'' where the response to a perturbation deviates from linear elasticity~\cite{ARCMP, Lerner2014, Hexner2018}. 
Thus, one might expect an approach based on simple local structural descriptors to decrease in performance as the pressure approaches zero, or at least for the number of necessary descriptors to increase strongly. 
Moreover, we find that $C_S$ shows no noticeable finite-size scaling at low pressures, 
further indicating that the slight decrease in accuracy at low pressures is unlikely due to a diverging correlation length~\cite{monster} (see Supporting Information~\cite{Supplement}).

    \begin{figure}[t!]
      \centering
     \includegraphics[width=\linewidth]{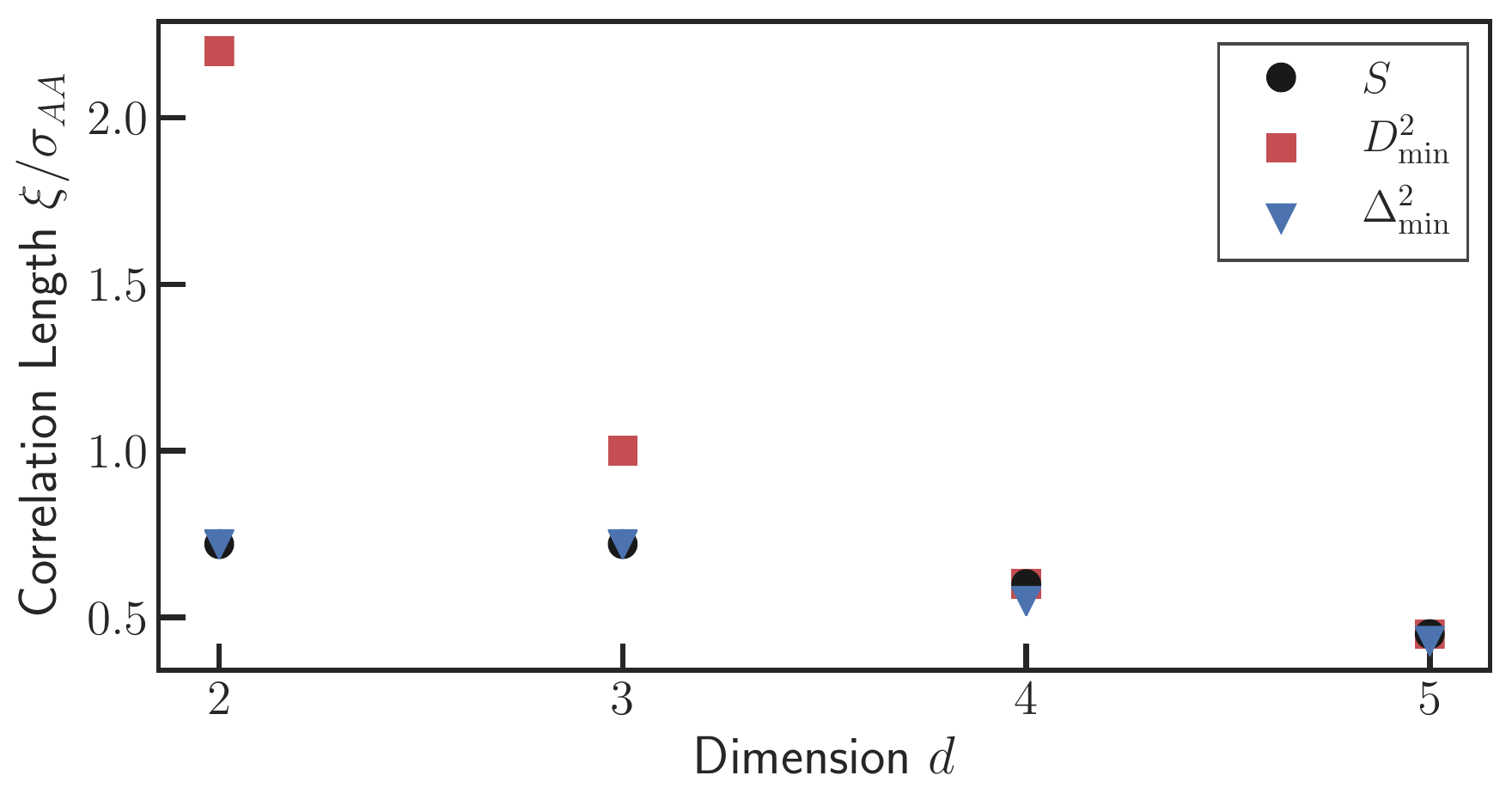}
      \caption{Correlation lengths extracted from fits to initial decay of small-small-particle correlation functions out to a distance of about $3$ particle diameters, calculated at the pressure $P^*(d)$ . All correlation lengths decrease slightly with increasing dimension $d$.  Red squares: $\dmin$, blue triangles: $\delmin$, black circles: $S$.} 
      \label{fig:corr}
      \end{figure}

%    \begin{figure}[t!]
%        \centering
%        \includegraphics[width=0.95\linewidth]{figure_Z_percentile.png}
%        \caption{The $Z$ percentile of the particle which has maximum $\dmin$ in a plastic event. }
%          \label{fig:Z}
%        \end{figure}

In contrast, it is not suprising that structural predictors other than $Z$ do not contribute to $S$ in higher $d$. 
In infinite dimensions, it suffices for an exact treatment to truncate the virial expansion at second order, i.e. to only include the effects of nearest interacting neighbors~\cite{parisi_theory_2020}. 
In high $d$, each particle's nearest neighbours are increasingly likely to also be neighbours of one another, so structural descriptors other than particle type, number of contacts, and the number of nearest-neighbour gaps should contain no information in the limit $d \to \infty$.

The most interesting aspect of our findings is that rearrangements continue to localize around low-$Z$ particles as the $d$ is increased, as evidenced by $C_\mathrm{Z}$.
Because the low-frequency vibrational at zero strain modes become more extended in higher $d$, one might expect the initial rearrangement would become less localized in higher dimensions, leading to a decrease of $C_\mathrm{Z}$~\cite{Charbonneau2016,Shimada2019}.
%The value of $C_S$ is not enough to determine how localized the initial rearrangement is as it measures only the rank of $S$ for the particle exhibiting the highest value of $\dmin$ in the lowest-frequency vibrational mode at the verge of a stress drop. 
To determine the extent of localziation of the initial rearrangement, 
we calculate the spatial correlations of $\dmin$ and $\delmin$, along with the spatial correlations of softness $S$. 
At long distances, the  $\dmin$ of each rearrangement displays power-law decay, consistent with continuum elasticity, but at short distances there is an exponential decay in $\dmin$ correlations on the scale of the particle diameter~\cite{Supplement}. 
Using this decay, we fit an exponential function to calculate an approximate correlation length. 
We do the same for the correlations in $\delmin$, for which small spatial gradients are scaled out, and for the softness $S$. 

Fig.~\ref{fig:corr} shows these correlation lengths evaluated at $P^*(d)$ as a function of $d$. We observe that the correlation lengths all decrease with dimension, indicating that rearrangements become more localized in space. Note, however, that there are more nearby particles with increasing $d$ so the rearrangements may be more extended in terms of the number of particles involved (e.g., higher participation ratio) even though they are more spatially localized. In the Supporting Information, we show that the structural correlation lengths, as well as that for $\delmin$, depend only weakly on pressure in each dimension, increasing slightly at higher pressures.
On the other hand, the $\dmin$ correlation length, grows as pressure is decreased as one might expect, although it does not seem to diverge at unjamming~\cite{Supplement}.
In the Supporting Information,  we also provide the inverse participation ratio of the critical modes for direct comparison with previous reports at zero strain~\cite{Charbonneau2016,Shimada2019}.   In our critical modes we do not see the same apparent delocalization in IPR which was reported at zero strain, although we do not hold N fixed.

      \begin{figure}[t!]
        \centering
       \includegraphics[width=\linewidth]{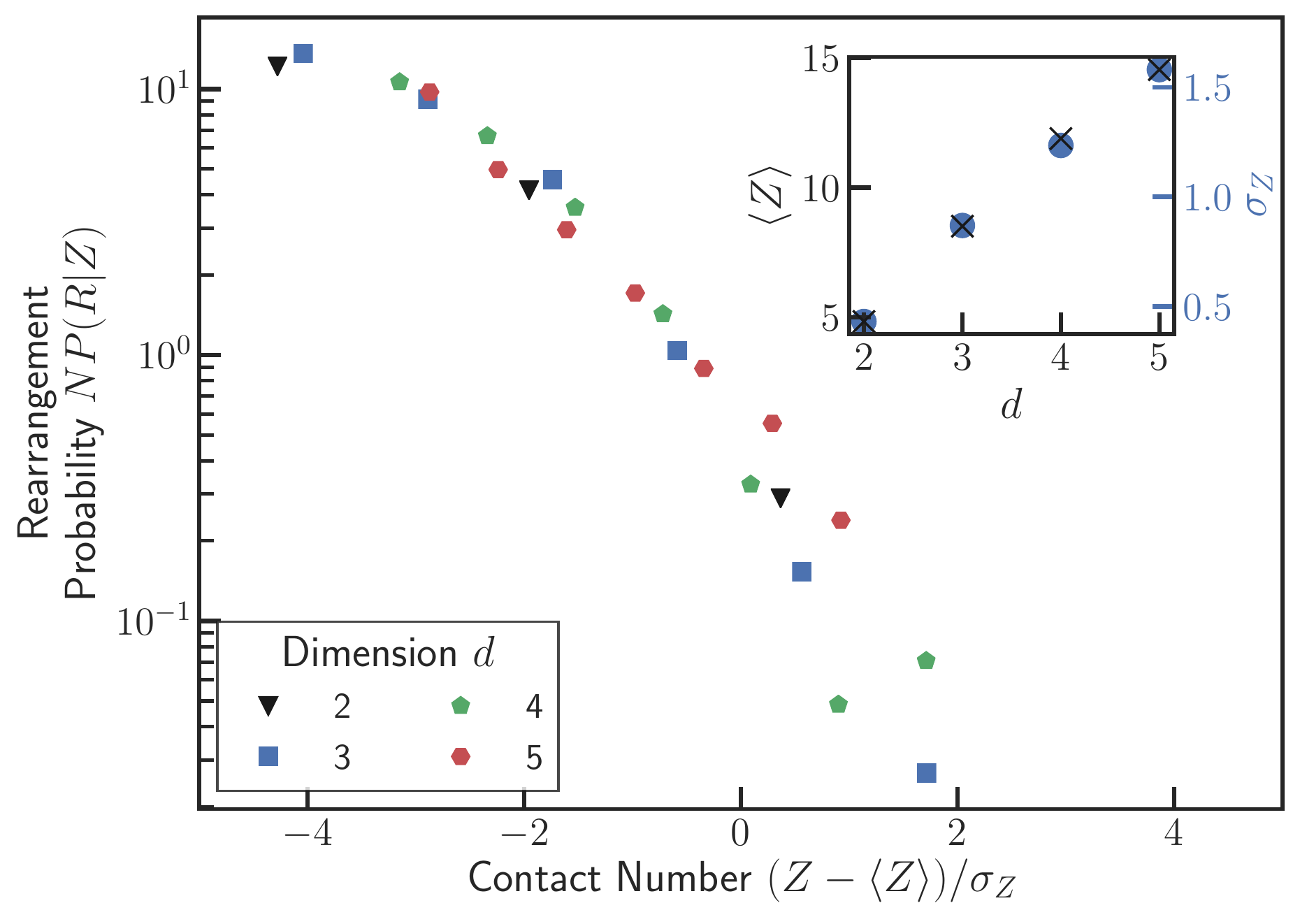}
        \caption{Probability of rearranging for small particles as a function of $(Z - \langle Z \rangle)/ \sigma_Z$ in varying spatial dimension ($d=2$ black triangles, $d=3$ blue squares, $d=4$  green pentagons, $d=5$  red hexagons). As dimension is varied, or pressure is varied (at pressures low enough for $Z$ to still be strongly predictive), the dependence of the rearrangement probability on this rescaled $Z$ is qualitatively similar, indicating that particles with relatively low $Z$ are always more likely to rearrange, even in higher $d$ where they are not bucklers. Inset shows $\langle Z\rangle$ (black x) and $\sigma_Z$ (blue circles) for small particles, indicating that at $P^*$ (where $\sigma_Z/\langle Z \rangle$ is smaller than at $P=0$) the asymptotic $1/\sqrt{d}$ scaling of $\sigma_Z / \langle Z \rangle$ has yet to occur.} 
        \label{fig:prz}
        \end{figure}

In Fig.~\ref{fig:corr}, we also observe that the spatial correlations of $\delmin$ are the same as those of $S$, indicating that structural correlations set the size of rearrangements. 
In earlier work~\cite{Cubuk2017}, it was concluded that the spatial correlations of $\dmin$ are the same as those of $S$; here we find that spatial correlations of $\dmin$ are noticeably longer in range in $d=2,3$. However, the systems studied in Ref.~\cite{Cubuk2017} did not exhibit elastic correlations over as long a range as we find here as a result of friction or thermal fluctuations. Therefore, it is likely that the correlation lengths of $\dmin$ and $\delmin$ in those systems would be far more similar.

It has previously been noted that ``bucklers'', particles with the minimum number of contacts for local stability, are associated with deviations from mean-field behaviour and become less and less common in higher-dimensional packings \cite{Charbonneau2015}. Naively, one might imagine that this would result in a decrease in $C_Z^*$ with dimension. 
To investigate this hypothesis, in Fig.~\ref{fig:prz} we report the probability of the rearrangement being located at a given particle as a function of $Z$ in each dimension at pressure $P^*(d)$.
We collapse the collapse the curves by multiplying probability by the total number of particles $N$ and standardizing $Z$ by subtracting off its mean over all particles $\expval*{Z}$ and dividing by its standard deviation $\sigma_Z$.
We find that particles that rearrange tend to have small $Z$ relative to the average, often by two or more standard deviations.
This result holds even as the $d$ increases and rattlers and bucklers become exceedingly rare,
demonstrating that it is a particle's value of $Z$ relative to the rest of the configuration that determines its propensity to rearrange, rather than solely its status as a buckler or a rattler.

In the inset to Fig.~\ref{fig:prz}, we show the scaling of $\langle Z \rangle$ and $\sigma_Z$ evaluated at $P^*(d)$ for each $d$ studied. 
We see that $\langle Z \rangle(P^*(d))$ scales linearly with $d$ as expected since $\langle Z \rangle(P=0)=2d$. However, we find $\sigma_Z(P^*(d)) \sim d$, while we would expect $\sigma_Z(P=0) \sim \sqrt{d}$ \cite{Charbonneau2015}.
 We find that $\sigma_Z/ \langle Z \rangle$ decreases as a function of $P$, and thus we would expect $\sigma_Z{\left(P^*\right)} \sim \sqrt{d}$ at large enough $d$ (since it is bounded above by $P=0$).
 Thus, although the collapse in Fig.~\ref{fig:prz} seems to indicate that lower $Z$ particles should dominate rearrangements in all $d$, there is the caveat that $\sigma_Z$ has not yet reached its asymptotic scaling in $d=5$.

In summary, our results suggest a refinement of the picture relating local structure to rearrangements in high dimensions. Local structure remains important in determining particle rearrangements. In fact, the spatial size of rearrangements, as measured by the decay of $\dmin$, is somewhat shorter in range. Nevertheless, the number of particles within that range increases exponentially with $d$, so that rearrangements may become extended in terms of the number of particles that participate. 
In high $d$, the contact number $Z$ for each particle becomes a good structural predictor of rearrangements, implying that rearrangements are controlled only by nearest neighbors that interact directly with a particle. Even though the contact number distribution narrows relative to the mean contact number $\langle Z \rangle \sim d$, the standard deviation diverges, scaling as $\sigma_Z \sim \sqrt{d}$~\cite{Charbonneau2015}. Thus, different particles can still have different $Z$ in high dimensions as $d \rightarrow \infty$, and low $Z$ particles may still have a higher propensity to rearrange. 

Defining $x = 2 Z/\langle Z \rangle \sim Z/d$, there are three simple scenarios that could prevail in high $d$: (1) $P_\mathrm{R}{\left(Z\right)} = f{\left(x\right)}$. (2) Just as the distribution of contacts has a large-deviation form $P{\left(Z\right)} \sim e^{-d B\left(x\right)}$, one might expect  $P_\mathrm{R}{\left(Z\right)} \sim e^{d A{\left(x\right)}}$ \cite{Charbonneau2015}. (3) The scenario of Fig.~\ref{fig:prz}, where $P_\mathrm{R}{\left(Z\right)} = f{\left((Z - \langle Z)\rangle  /\sigma_Z\right)}$. In the first scenario, the typical $x^*$ for rearranging particles goes to $1$ in and $C_Z \to \frac{1}{2}$ (no correlation) as $d \rightarrow \infty$.  In the second, $1< x^* < 2$: rearranging particles in large dimension are not bucklers, but are still extremely atypical, and $C_\mathrm{Z} \to 1$. In the third scenario, $x^* \to 2$ as $d\to \infty$, but $2 - x^*$ goes to zero at the same rate as $\sigma_\mathrm{Z}$, so rearranging particles are still atypical and $C_{\mathrm{Z}}$ is constant. We find the large deviation form of scenario (2) does not collapse our data as well as in Fig.~\ref{fig:prz}, but this could be because $d=5$ is still low.

Either scenario (2) or (3) would connect the softness picture of dynamical behavior associated with the glass transition in $d=3$ with mean field theory.

\section*{Acknowledgements}

We thank Ge Zhang, Sam Schoenholz, Robert Ivancic, Fran\c{c}ois Landes, Eric Corwin and Francesco Zamponi for helpful discussions. We thank Carl Goodrich for providing useful code.

This research was supported by the US Department of Energy, Office of Basic Energy Sciences,
Division of Materials Sciences and Engineering under Award DE-FG02-05ER46199 (J.W.R), 
the Natural Sciences and Engineering Research Council of Canada under a Postgraduate Scholarship -- Doctoral award (S.A.R), 
and the Simons Foundation for the collaboration
Cracking the Glass Problem via award 454945 (J.W.R, S.A.R, and A.J.L) and Investigator Award 327939 (A.J.L.).

\bibliography{ms}

\begin{thebibliography}{10}

\bibitem{Reichman2008}
Asaph Widmer-Cooper, Heidi Perry, Peter Harrowell, and David~R. Reichman.
\newblock {Irreversible reorganization in a supercooled liquid originates from
  localized soft modes}.
\newblock {\em Nature Physics}, 4(9):711--715, 2008.

\bibitem{Reichman2009}
Asaph Widmer-Cooper, Heidi Perry, Peter Harrowell, and David~R. Reichman.
\newblock Localized soft modes and the supercooled liquid’s irreversible
  passage through its configuration space.
\newblock {\em The Journal of Chemical Physics}, 131(19):194508, 2009.

\bibitem{Manning2011}
M.~L. Manning and A.~J. Liu.
\newblock {Vibrational modes identify soft spots in a sheared disordered
  packing}.
\newblock {\em Physical Review Letters}, 107(10):108302, aug 2011.

\bibitem{SamPRX}
S.~S. Schoenholz, A.~J. Liu, R.~A. Riggleman, and J.~Rottler.
\newblock Understanding plastic deformation in thermal glasses from
  single-soft-spot dynamics.
\newblock {\em Phys. Rev. X}, 4:031014, Jul 2014.

\bibitem{barrat}
Michel Tsamados, Anne Tanguy, Chay Goldenberg, and Jean-Louis Barrat.
\newblock Local elasticity map and plasticity in a model lennard-jones glass.
\newblock {\em Phys. Rev. E}, 80:026112, Aug 2009.

\bibitem{Gartner2016}
Luka Gartner and Edan Lerner.
\newblock Nonlinear plastic modes in disordered solids.
\newblock {\em Phys. Rev. E}, 93:011001, Jan 2016.

\bibitem{Zylberg2017}
Jacques Zylberg, Edan Lerner, Yohai Bar-Sinai, and Eran Bouchbinder.
\newblock Local thermal energy as a structural indicator in glasses.
\newblock {\em Proceedings of the National Academy of Sciences},
  114(28):7289--7294, 2017.

\bibitem{malins}
Alex Malins, Jens Eggers, {C. Patrick} Royall, {Stephen R.} Williams, and
  Hajime Tanaka.
\newblock Identification of long-lived clusters and their link to slow dynamics
  in a model glass former.
\newblock {\em Journal of Chemical Physics}, 138(12), March 2013.

\bibitem{Tanaka2018}
Hua Tong and Hajime Tanaka.
\newblock Revealing hidden structural order controlling both fast and slow
  glassy dynamics in supercooled liquids.
\newblock {\em Phys. Rev. X}, 8:011041, Mar 2018.

\bibitem{Richard2020}
D.~Richard, M.~Ozawa, S.~Patinet, E.~Stanifer, B.~Shang, S.~A. Ridout, B.~Xu,
  G.~Zhang, P.~K. Morse, J.~L. Barrat, L.~Berthier, M.~L. Falk, P.~Guan, A.~J.
  Liu, K.~Martens, S.~Sastry, D.~Vandembroucq, E.~Lerner, and M.~L. Manning.
\newblock Predicting plasticity in disordered solids from structural
  indicators, 2020.

\bibitem{tetrahedrality}
Susana Mar\'{\i}n-Aguilar, Henricus~H. Wensink, Giuseppe Foffi, and Frank
  Smallenburg.
\newblock Tetrahedrality dictates dynamics in hard sphere mixtures.
\newblock {\em Phys. Rev. Lett.}, 124:208005, May 2020.

\bibitem{Bapst2020}
V.~Bapst, T.~Keck, A.~Grabska-Barwi{\'{n}}ska, C.~Donner, E.~D. Cubuk, S.~S.
  Schoenholz, A.~Obika, A.~W.R. Nelson, T.~Back, D.~Hassabis, and P.~Kohli.
\newblock {Unveiling the predictive power of static structure in glassy
  systems}.
\newblock {\em Nature Physics}, 16(4):448--454, 2020.

\bibitem{boattini2020autonomously}
Emanuele Boattini, Susana Marín-Aguilar, Saheli Mitra, Giuseppe Foffi, Frank
  Smallenburg, and Laura Filion.
\newblock Autonomously revealing hidden local structures in supercooled
  liquids, 2020.

\bibitem{Viitanen2020}
Leevi Viitanen, Jonatan R.~Mac Intyre, Juha Koivisto, Antti Puisto, and Mikko
  Alava.
\newblock Machine learning and predicting the time dependent dynamics of local
  yielding in dry foams, 2020.

\bibitem{Cubuk2015}
E.~D. Cubuk, S.~S. Schoenholz, J.~M. Rieser, B.~D. Malone, J.~Rottler, D.~J.
  Durian, E.~Kaxiras, and A.~J. Liu.
\newblock {Identifying structural flow defects in disordered solids using
  machine-learning methods}.
\newblock {\em Physical Review Letters}, 114(10):108001, mar 2015.

\bibitem{Schoenholz2016}
S.~S. Schoenholz, E.~D. Cubuk, D.~M. Sussman, E.~Kaxiras, and A.~J. Liu.
\newblock {A structural approach to relaxation in glassy liquids}.
\newblock {\em Nature Physics}, 12(5):469--471, feb 2016.

\bibitem{Sussman2017}
Daniel~M. Sussman, Samuel~S. Schoenholz, Ekin~D. Cubuk, and Andrea~J. Liu.
\newblock {Disconnecting structure and dynamics in glassy thin films}.
\newblock {\em Proceedings of the National Academy of Sciences of the United
  States of America}, 114(40):10601--10605, 2017.

\bibitem{Sharp2018}
Tristan~A. Sharp, Spencer~L. Thomas, Ekin~D. Cubuk, Samuel~S. Schoenholz,
  David~J. Srolovitz, and Andrea~J. Liu.
\newblock {Machine learning determination of atomic dynamics at grain
  boundaries}.
\newblock {\em Proceedings of the National Academy of Sciences of the United
  States of America}, 115(43):10943--10947, 2018.

\bibitem{Harrington2018}
Matt Harrington and Douglas~J. Durian.
\newblock Anisotropic particles strengthen granular pillars under compression.
\newblock {\em Phys. Rev. E}, 97:012904, Jan 2018.

\bibitem{Ma2019}
Xiaoguang Ma, Zoey~S. Davidson, Tim Still, Robert J.~S. Ivancic, S.~S.
  Schoenholz, A.~J. Liu, and A.~G. Yodh.
\newblock Heterogeneous activation, local structure, and softness in
  supercooled colloidal liquids.
\newblock {\em Phys. Rev. Lett.}, 122:028001, Jan 2019.

\bibitem{Landes2020}
Fran{\c{c}}ois~P. Landes, Giulio Biroli, Olivier Dauchot, Andrea~J. Liu, and
  David~R. Reichman.
\newblock {Attractive versus truncated repulsive supercooled liquids: The
  dynamics is encoded in the pair correlation function}.
\newblock {\em Physical Review E}, 101(1):1--5, 2020.

\bibitem{Cubuk2020}
Ekin~D. Cubuk, Andrea~J. Liu, Efthimios Kaxiras, and Samuel~S. Schoenholz.
\newblock Unifying framework for strong and fragile liquids via machine
  learning: a study of liquid silica, 2020.

\bibitem{Tah2020}
Indrajit Tah, Tristan~A. Sharp, Andrea~J. Liu, and Daniel~M. Sussman.
\newblock Quantifying the link between local structure and cellular
  rearrangements using information in models of biological tissues, 2020.

\bibitem{Zhang2020}
Ge~Zhang, Sean Ridout, and Andrea~J. Liu.
\newblock Interplay of rearrangements, strain, and local structure during
  avalanche propagation, 2020.

\bibitem{Rocks2019}
Jason~W. Rocks, Sean~A. Ridout, and Andrea~J Liu.
\newblock {Learning-based approach to plasticity in athermal sheared amorphous
  packings: Improving softness}, 2020.

\bibitem{Charbonneau2017}
Patrick Charbonneau, Jorge Kurchan, Giorgio Parisi, Pierfrancesco Urbani, and
  Francesco Zamponi.
\newblock {Glass and Jamming Transitions: From Exact Results to
  Finite-Dimensional Descriptions}.
\newblock {\em Annual Review of Condensed Matter Physics}, 8(1):265--288, 2017.

\bibitem{Berthier2019}
Ludovic Berthier, Giulio Biroli, Patrick Charbonneau, Eric~I. Corwin, Silvio
  Franz, and Francesco Zamponi.
\newblock {Gardner physics in amorphous solids and beyond}.
\newblock {\em Journal of Chemical Physics}, 151(1), 2019.

\bibitem{Charbonneau2012}
Patrick Charbonneau, Eric~I. Corwin, Giorgio Parisi, and Francesco Zamponi.
\newblock {Universal microstructure and mechanical stability of jammed
  packings}.
\newblock {\em Physical Review Letters}, 109(20):1--5, 2012.

\bibitem{Charbonneau2015}
Patrick Charbonneau, Eric~I. Corwin, Giorgio Parisi, and Francesco Zamponi.
\newblock {Jamming criticality revealed by removing localized buckling
  excitations}.
\newblock {\em Physical Review Letters}, 114(12):1--5, 2015.

\bibitem{Charbonneau2016}
Patrick Charbonneau, Eric~I. Corwin, Giorgio Parisi, Alexis Poncet, and
  Francesco Zamponi.
\newblock Universal non-debye scaling in the density of states of amorphous
  solids.
\newblock {\em Phys. Rev. Lett.}, 117:045503, Jul 2016.

\bibitem{Shimada2019}
Masanari Shimada, Hideyuki Mizuno, Ludovic Berthier, and Atsushi Ikeda.
\newblock Low-frequency vibrations of jammed packings in large spatial
  dimensions, 2019.

\bibitem{vanHecke2014}
Merlijn~S. van Deen, Johannes Simon, Zorana Zeravcic, Simon Dagois-Bohy,
  Brian~P. Tighe, and Martin van Hecke.
\newblock Contact changes near jamming.
\newblock {\em Phys. Rev. E}, 90:020202, Aug 2014.

\bibitem{Morse2020}
Peter Morse, Sven Wijtmans, Merlijn van Deen, Martin van Hecke, and M.~Lisa
  Manning.
\newblock Differences in plasticity between hard and soft spheres.
\newblock {\em Phys. Rev. Research}, 2:023179, May 2020.

\bibitem{Tewari1999}
Shubha Tewari, Dylan Schiemann, Douglas~J. Durian, Charles~M. Knobler,
  Stephen~A. Langer, and Andrea~J. Liu.
\newblock {Statistics of shear-induced rearrangements in a two-dimensional
  model foam}.
\newblock {\em Physical Review E - Statistical Physics, Plasmas, Fluids, and
  Related Interdisciplinary Topics}, 60(4):4385--4396, 1999.

\bibitem{Maloney2006}
Craig~E. Maloney and Ana{\"{e}}l Lema{\^{i}}tre.
\newblock {Amorphous systems in athermal, quasistatic shear}.
\newblock {\em Physical Review E - Statistical, Nonlinear, and Soft Matter
  Physics}, 74(1):1--22, 2006.

\bibitem{Falk1998}
M~L Falk and J~S Langer.
\newblock {Dynamics of viscoplastic deformation in amorphous solids M.}
\newblock {\em Physical Review E}, 57(6):14, 1998.

\bibitem{Supplement}
See Supplemental Material at [URL will be inserted by publisher] for data
  showing lack of finite size effects, regression $R^2$ as a function of
  pressure, and correlation lengths and $P{\left(R|Z\right)}$ as a function of
  pressure..

\bibitem{Patinet2016}
Sylvain Patinet, Damien Vandembroucq, and Michael~L. Falk.
\newblock {Connecting Local Yield Stresses with Plastic Activity in Amorphous
  Solids}.
\newblock {\em Physical Review Letters}, 117(4):045501, jul 2016.

\bibitem{ARCMP}
Andrea~J. Liu and Sidney~R. Nagel.
\newblock {The jamming transition and the marginally jammed solid}.
\newblock {\em Annual Review of Condensed Matter Physics}, 1:347--369, 2010.

\bibitem{Lerner2014}
Edan Lerner, Eric Degiuli, Gustavo D{\"{u}}ring, and Matthieu Wyart.
\newblock {Breakdown of continuum elasticity in amorphous solids}.
\newblock {\em Soft Matter}, 10(28):5085--5092, 2014.

\bibitem{Hexner2018}
Daniel Hexner, Andrea~J. Liu, and Sidney~R. Nagel.
\newblock {Two Diverging Length Scales in the Structure of Jammed Packings}.
\newblock {\em Physical Review Letters}, 121(11):115501, 2018.

\bibitem{monster}
Carl~P. Goodrich, Simon Dagois-Bohy, Brian~P. Tighe, Martin van Hecke,
  Andrea~J. Liu, and Sidney~R. Nagel.
\newblock Jamming in finite systems: Stability, anisotropy, fluctuations, and
  scaling.
\newblock {\em Phys. Rev. E}, 90:022138, Aug 2014.

\bibitem{parisi_theory_2020}
Giorgio Parisi, Pierfrancesco Urbani, and Francesco Zamponi.
\newblock {\em Theory of {Simple} {Glasses}: {Exact} {Solutions} in {Infinite}
  {Dimensions}}.
\newblock Cambridge University Press, New York, February 2020.

\bibitem{Cubuk2017}
E.~D. Cubuk, R.~J.~S. Ivancic, S.~S. Schoenholz, D.~J. Strickland, A.~Basu,
  Z.~S. Davidson, J.~Fontaine, J.~L. Hor, Y.-R. Huang, Y.~Jiang, N.~C. Keim,
  K.~D. Koshigan, J.~A. Lefever, T.~Liu, X.-G. Ma, D.~J. Magagnosc, E.~Morrow,
  C.~P. Ortiz, J.~M. Rieser, A.~Shavit, T.~Still, Y.~Xu, Y.~Zhang, K.~N.
  Nordstrom, P.~E. Arratia, R.~W. Carpick, D.~J. Durian, Z.~Fakhraai, D.~J.
  Jerolmack, Daeyeon Lee, Ju~Li, R.~Riggleman, K.~T. Turner, A.~G. Yodh, D.~S.
  Gianola, and Andrea~J. Liu.
\newblock Structure-property relationships from universal signatures of
  plasticity in disordered solids.
\newblock {\em Science}, 358(6366):1033--1037, 2017.

\end{thebibliography}

%%%%%%%%%% Merge with supplemental materials %%%%%%%%%%
\pagebreak
\widetext
\begin{center}
\textbf{\large Supporing Information:\\ Correlation of plastic events with local structure in jammed packings across spatial dimensions}
\end{center}
%%%%%%%%%% Merge with supplemental materials %%%%%%%%%%
%%%%%%%%%% Prefix a "S" to all equations, figures, tables and reset the counter %%%%%%%%%%
\setcounter{equation}{0}
\setcounter{figure}{0}
\setcounter{table}{0}
\setcounter{page}{1}
\makeatletter

\renewcommand{\thefigure}{S\arabic{figure}}

\begin{figure}[h!]
  \centering
 \includegraphics[width=0.5\linewidth]{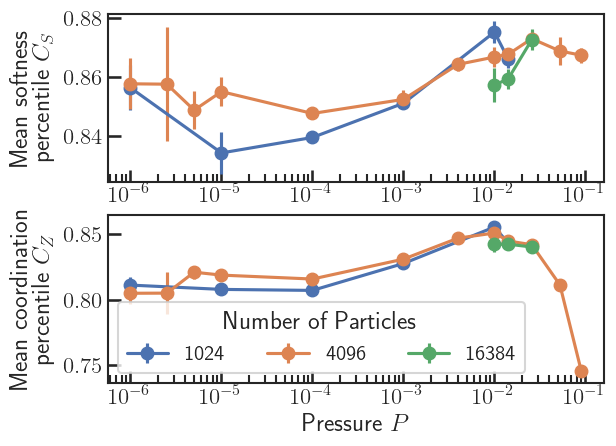}
  \caption{$C$ for various $N$ in $2d$. This data shows no evidence of the finite-size scaling seen in other properties associated with a diverging correlation length at $P=0$, i.e., the low pressure at which accuracy decreases does not become higher at smaller $N$.
  } 
  \label{fig:SIss}
  \end{figure}

\begin{figure}[h!]
  \centering
 \includegraphics[width=0.5\linewidth]{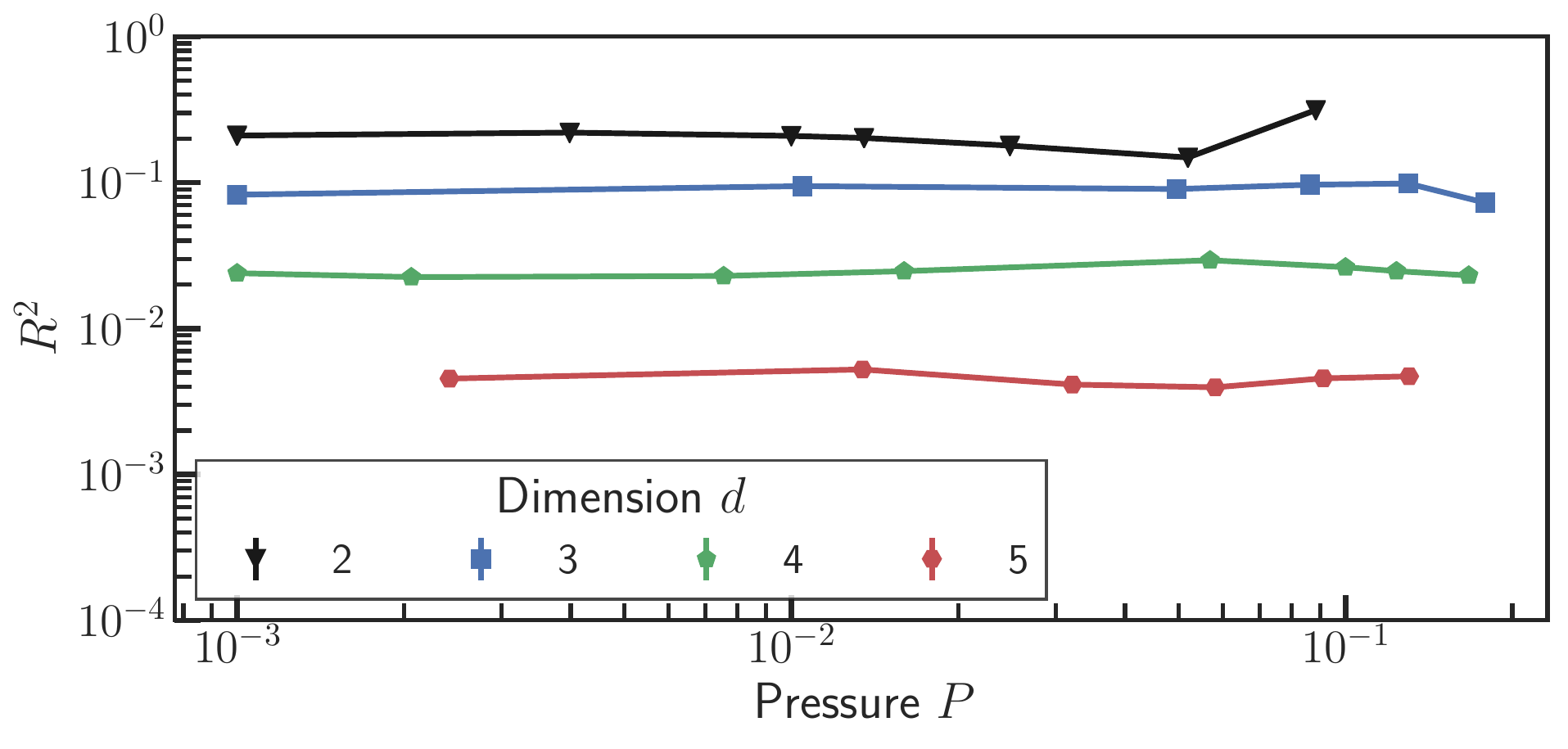}
  \caption{$R^2$ for regression against $\Delta^2_{\mathrm{min}}$ as defined in main text.  In addition to the decay with dimension shown in the main text, here the very weak pressure dependence is made clear.} 
  \label{fig:SIr2}
  \end{figure}

\begin{figure}[h!]
\centering
\includegraphics[width=0.5\linewidth]{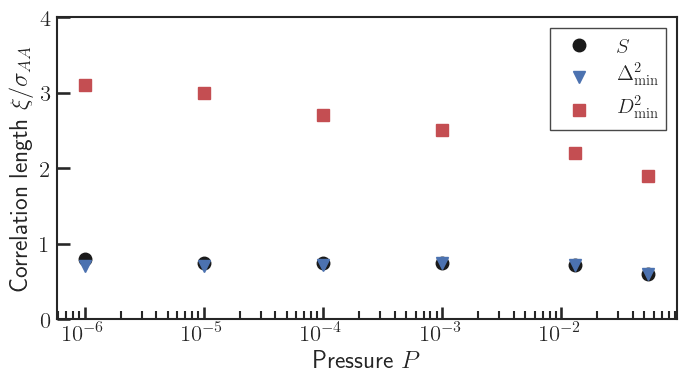}
\caption{Correlation lengths extracted from fits to initial decay of small-small-particle correlation functions in $2d$, showing weak dependence on pressure. The $\dmin$ correlation length depends more on pressure, but does not seem to diverge at unjamming.  Blue circles: $\dmin$, orange squares: $\Delta^2_{\mathrm{min}}$, green triangles: $S$.} 
\label{fig:SIcorrP}
\end{figure}

\begin{figure}[h!]
\centering
\includegraphics[width=0.5\linewidth]{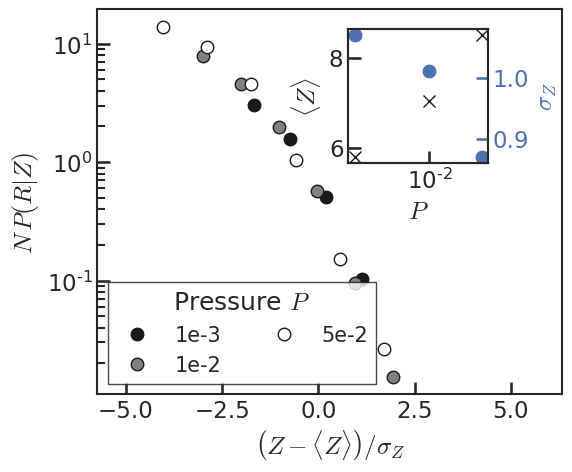}
\caption{Probability of rearranging for small particles as a function of $(Z - \langle Z \rangle)/ \sigma_Z$ in several pressures in $3d$. ($P=1e-3$ black, $P=1e-2$ gray, $P=5e-2$ white). Leaving out pressures above $P^*$, the dependence of the rearrangement probability on this rescaled $Z$ is qualitatively similar at different $P$. Inset shows $\langle Z\rangle$ (black x's) and $\sigma_Z$ (blue circles) for small particles. Since $\langle Z \rangle$ and $\sigma_Z$ are anticorrelated, rather than correlated as when $d$ is varied in the main text, this provides further evidence that the rescaling used is the correct one. } 
\label{fig:SIprz}
\end{figure}

\begin{figure}[h!]
\centering
\includegraphics[width=0.5\linewidth]{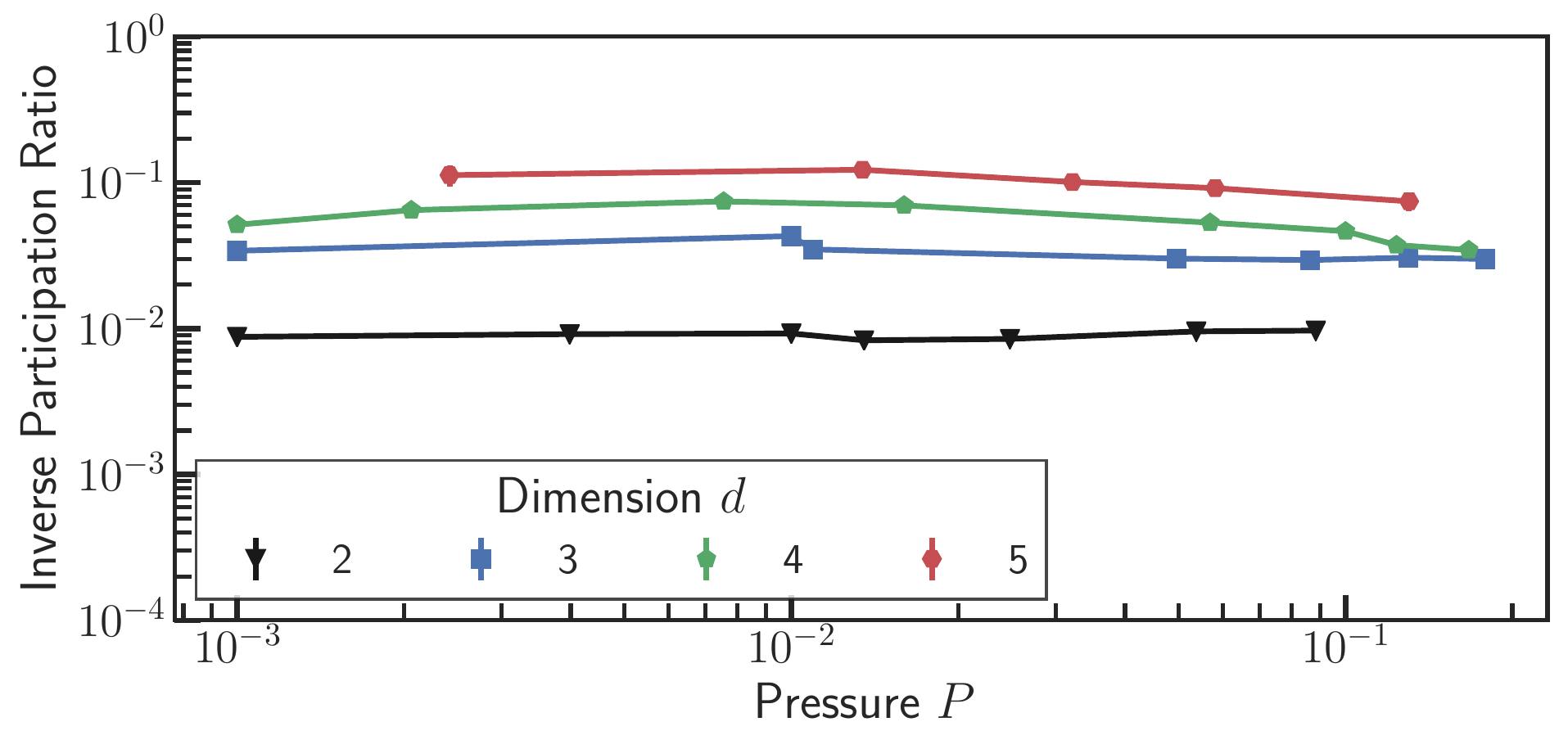}
\caption{ Inverse participation ratio (IPR) $\sum_{i=1}^{N} u_i^4 / \sum_{i=1}^N u_i^2$ of critical modes (with displacement $\mathbf{u}$). Black triangles: $2d$, $N=4096$. Blue squares: $3d$, $N=8192$. Green pentagons: $4d$, $N=8192$. Red hexagons: $5d$, $N=16384$. Because it is computationally intensive to study larger systems in high $d$, we did not look at the dependence of the participation ratio on system size, which is needed in order to determine whether the modes are spatially localized or not } 
\label{fig:SIipr}
\end{figure}

\end{document}